# Demonstration of a 3rd order hierarchy of higher order topological states in a three-dimensional acoustic metamaterial

Matthew Weiner[1,2], Xiang Ni[1,2], Mengyao Li[1,2], Andrea Alù[3,2,1], and Alexander B. Khanikaev[1,2]

[1]Department of Electrical Engineering, Grove School of Engineering, City College of the City University of New York, 140th Street and Convent Avenue, New York, NY 10031, USA.
[2]Physics Program, Graduate Center of the City University of New York, New York, NY 10016, USA.
[3]Photonics Initiative, Advanced Science Research Center, City University of New York, New York, NY 10031, USA

**Abstract:** In the past years classical wave-systems have constituted an excellent platform for emulating complex quantum phenomena. This approach has been especially fruitful in demonstrating topological phenomena in photonics and acoustics: from chiral edge states of Chern insulators and helical edge states of topological insulators to higher-dimensional topological states of quasiperiodic systems and systems with synthetic dimensions. Recently, a new class of topological states localized in more than one dimension of a *D*-dimensional system, referred to as higher-order topological (HOT) states, has been reported, offering an even more versatile platform to confine and control classical radiation and mechanical motion. However, because experimental research of HOT states has so far been limited to two-dimensional (2D) systems, third and higher-order states have evaded experimental observation. Studying higher-dimensional classical systems therefore opens an opportunity to emulate higher-order topological insulators and explore HOT states beyond second order. In this letter, we design and experimentally study a 3D acoustic metamaterial supporting third-order (0D) topological corner states along with second-order (1D) edge states within the same topological bandgap, thus establishing a full hierarchy of HOT states in three dimensions. The metamaterial is implemented over a versatile additive manufacturing platform, which enables rapid prototyping of metaatoms and metamolecules, which can be snapped together to form 3D metamaterials with complex geometries. The assembled 3D topological metamaterial represents the acoustic analogue of a pyrochlore lattice made of interconnected molecules, and is shown to exhibit topological bulk polarization, leading to the emergence of HOT states localized in all three or in two dimensions.

Topological acoustics and mechanics have been explored as a powerful platform for the implementation of a plethora of topological phenomena (*1-4*). A bias for sound propagation imparted by the angular-momentum carried by a rotating fluid or by rotational motion in mechanical resonators has been used to emulate the effect of magnetic bias and to demonstrate the emergence of quantum Hall effect (QHE)-like states with robust edge transport for acoustics and mechanics (*5-11*). In parallel, a variety of symmetry-protected topological acoustic states have been reported, including Zak-phase in 1D acoustic lattices (*12*), quantum spin Hall effect (QSHE) (*13-16*), nontrivial bulk polarization induced edge and corner states in 2D Kagome lattices (*17-20*). More recently, versatile 3D structures fabricated by either direct assembly or by advanced 3D manufacturing have been used to demonstrate Weyl points and surface Fermi arc-like states in 3D acoustic and mechanical metamaterials (*21-24*) as well as mechanical quadrupolar topological insulators (*25-27*). Indeed, 3D geometries dramatically expand the design opportunities for acoustic topological metamaterials, thus offering the possibility of demonstration of unprecedented topological phenomena, often still unobserved in their condensed matter or photonic analogues. High-order topological insulators with a hierarchy of states represent one example in which the system dimensionality becomes crucial. HOT states have so far been demonstrated only in 2D topological systems, being limited therefore to second-order states (*17-19, 28*). Higher-order hierarchies of topological states have so far evaded experimental observation. In this work, we experimentally observe for the first time such hierarchy of second- and third-order states in a 3D acoustic metamaterial. We exploit 3D printing and a snap assembly technique to construct an acoustic analogue of the topological pyrochlore lattice, which hosts a third-order hierarchy of topological states. We show that the manufactured acoustic metacrystal possesses topological bulk polarization, and we experimentally demonstrate for the first time a hierarchy of HOT states in three-dimensional metamaterials by observing both $2^{nd}$-order edge and $3^{rd}$-order corner states localized in two and three dimensions, respectively.

**Topological acoustic metamaterial design**

A three-dimensional acoustic metamaterial emulator of the pyrochlore lattice (*28, 29*) is realized using meta-molecules (unit cells) schematically shown in Fig 1A and B, whose specific dimensions are given in the Methods section. Each molecule consists of four acoustic resonators, with acoustic pressure modes oscillating along the axial direction. We choose to work with the fundamental mode, whose only node appears in the center of the resonator. To form the lattice, the resonators are coupled through narrow cylindrical channels whose position on the resonator is carefully chosen to respect the crystalline symmetries of the pyrochlore lattice. The control over coupling between resonators within and between adjacent metamolecules of the metacrystal, which is required to induce bulk topological polarization, is achieved by varying the diameter of the cylindrical coupling channels. Since the acoustic modes are strongly bound to the resonant cavities and the design of the structure only allows nearest neighbor coupling, the tight-binding model (TBM) can be an effective modelling tool for our system, with nearest-neighbor hopping described by intra-cell $\kappa$ and inter-cell $\gamma$ coupling parameters. COMSOL Multiphysics (Acoustic Module) has been used to verify the results with full-wave finite element method (FEM) simulations. If the diameter of the intra-connected channels is smaller than the one of the inter-connected channels, $\kappa < \gamma$, the structure is referred to as the "expanded" crystal, implying stronger coupling among resonators of the adjacent metamolecules. In the opposite condition, $\kappa > \gamma$, the structure is "shrunken", and modes are strongly coupled within the resonators of the metamolecule within the same unit cell. The acoustic resonators in the metamaterial are aligned to show the sites and primitive vectors of the face center cubic (FCC) lattice in Fig. 1B, and the corresponding first Brillouin zone shown in Fig 1C. As the dimensions of the crystal exceeded the print volume of currently available STL 3D printers, the hybrid approach, combining 3D printing of the constitutive elements and subsequent snap-assembly, was used. The metamolecules were 3D printed and snapped together using the interlocking features that are deliberately introduced in the design, thus allowing for the assembly of rigid and stable large scale metacrystals. The assembled finite structure consisting of 20 metamolecules (80 resonators) is shown in Fig. 1D. Note that the boundaries of the crystal are terminated by smaller resonators, which are designed to have resonant frequencies detuned from the spectral region of interest to high frequency range, yielding an effective boundary condition analogous to the terminated boundary in TBM.

**Non-trivial topology and bulk polarization**

First, we theoretically explore the band structure of the unit cell, as shown in Fig. 2A and B. In the ideal pyrochlore lattice, with $\kappa = \gamma$ (Fig. 2A), a line of degeneracy along the X-W high symmetry path can be observed. The degeneracy of this line is broken by making the inter-cell and intra-cell couplings different $\kappa \neq \gamma$ (Fig. 2B). Through both semi-analytical TBM and the first-principles FEM modeling of the structure, we observe band inversion at the W point between two cases of expanded crystal and shrunken crystal, which indicates a nontrivial topological transition. However, energy bands alone are not sufficient to determine the topological difference between expanded and shrunken crystals, and further analysis of the associated eigenstates is required.

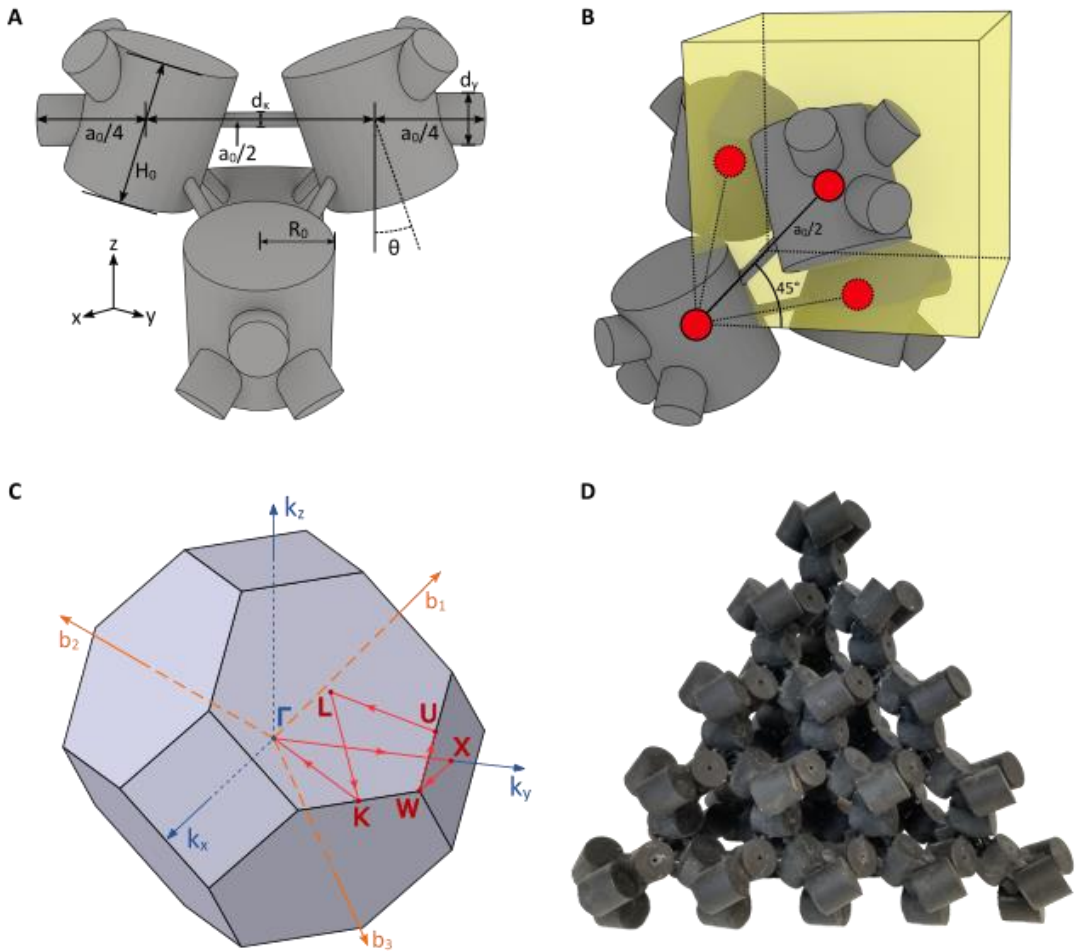

**Fig. 1 Concept and design of the acoustic pyrochlore lattice.** (**A**) Schematic and realistic design of the Wigner–Seitz unit cell of the expanded pyrochlore lattice. (**B**) Schematic of the Wigner–Seitz unit cell

inserted into the FCC lattice structure. (**C**) The first Brillouin zone of the FCC lattice, with labeled high symmetry points, high symmetry paths, and reciprocal lattice vectors. (**D**) Photograph of the 3D topological metamaterial assembled from 3D printed meta-molecules, with boundary cells attached. Each side length of the crystal consists of 4 meta-molecules (12 acoustic resonators).

In order to explicitly confirm the topological phase transition between expanded and shrunken crystals, we calculated the bulk topological polarization using the eigenvalue problem of the Wilson loops (*30*). The bulk polarization of the lowest energy band, defining the emergence of HOT states in the first band gap above this band for both cases of the expanded and shrunken lattices, can be obtained through the Wannier band $v_3(k_1, k_2)$, where $(k_1, k_2)$ vary along the reciprocal lattice vectors $(b_1, b_2)$ in momentum space, and the Wilson loop is evaluated along the direction of the reciprocal lattice vector $\boldsymbol{b_3}$ (Fig 1C). Then, the bulk polarization of the lowest energy band of interest is defined as

$$p_3(k_1, k_2) = \frac{1}{N_1 N_2} \sum_{b_3} v_3(k_1, k_2), \tag{1}$$

where $N_{1,2}$ is the total number of points in the discretized momentum space. Due to the spatial symmetries of the pyrochlore crystal, the bulk polarizations with cyclically permutated indexes are equivalent to each other, and therefore the system is characterized by nontrivial bulk polarization in all three dimensions. As shown in Fig 2C and D, the bulk polarization of the expanded crystal calculated through Eq. (1) yields ¼, which indicates that the average position of the modes belonging to the lowest energy band has a $\frac{a_0}{4}$ deviation (shift) from the center of the sites in the unit cell. According to the bulk-boundary correspondence principle (*19*), when boundaries are introduced in the expanded pyrochlore lattice with nonvanishing bulk polarization, the boundary modes such as surface, edge and corner states should appear hanging at the dangling sites of the corresponding terminations of the crystal. Meanwhile, the bulk polarization of the shrunken crystal is zero, implying that the average position of the desired modes resides at the center of the four sites in the metamolecule, and no boundary modes are present at any boundary.

**Experimental demonstration of a hierarchy of higher-order topological states**

We numerically calculated the energy spectrum of the metamaterial sample in Fig. 1d as a function of the ratio $\kappa/\gamma$ (Fig. 3A) by using TBM of the finite crystal. Based on the prediction from nontrivial bulk polarization outlined in the previous section, for the case of expanded lattice (when $\frac{\kappa}{\gamma} < 1$) we expect surface modes (purpled colored), edge modes (blue colored) and corner modes (red colored) to emerge at the boundaries. Both TBM and numerical simulations based on finite element method (FEM) fully confirm these predictions. Interestingly, when $\frac{\kappa}{\gamma} < \frac{1}{3}$, 4-fold degeneracy corner modes split from the bulk continuum (yellow colored bands), and, as can be proven analytically (see S.M. for details), the transition point indeed occurs at $\frac{\kappa}{\gamma} = \frac{1}{3}$.

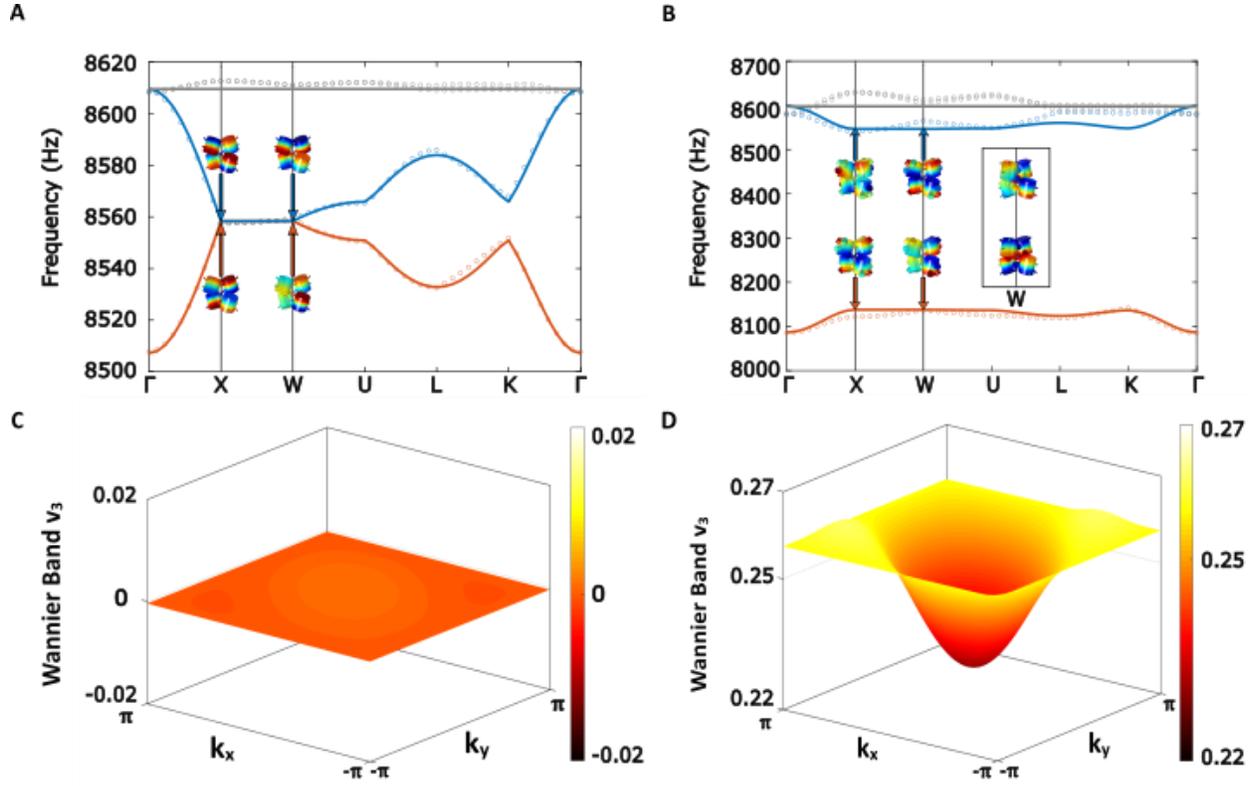

**Fig. 2 Theoretical prediction of nontrivial bulk polarization.** (**A**) Dispersion relation of the un-dimerized and (**B**) dimerized (expanded or shrunken) unit cell obtained by TBM (solid lines) and first-principle COMSOL calculations (circular dots) along a high symmetry path through the Brillouin zone shown in Fig 1C, and the corresponding mode profiles shown at X and W points. The inset in (**B**) shows the mode profiles of the shrunken unit cell at X and W, demonstrating the characteristic band inversion of topological systems. (**C,D**) Wannier bands of the pyrochlore lattice obtained by TBM for (**C**) the shrunken and (**D**) expanded unit cell. The ratio of intra-cell and inter-cell hopping terms are $\kappa/\gamma = 0.111$ and $\kappa/\gamma = 9.009$ for the expanded and shrunken cases, respectively.

In the 3D printed structure fabricated for our experimental investigation, we choose to work in the $\kappa/\gamma = 0.111$ regime of the energy spectrum of the lattice (indicated by dashed line in Fig.3A). In this case, the zero-energy corner states, as well as the edge and surface states, appear fully immersed into the bulk band gap and show no overlap with the bulk continuum. In agreement with the TBM energy spectrum, the density of states extracted from acoustic measurements on the structure (Fig. 3B) reveals a wide bulk band gap, as well as a hierarchy of topological boundary modes, including surface states, and HOT edge states and (four-fold degenerate) corner states (localized at the four corners of the structure). The Methods section provides details of the procedure to extract the density of states from experimental measurements. Interestingly, the corner states appear at "zero-energy", i.e. their frequency corresponds to the frequency of a separate individual resonator, however, the collective nature of these modes is evidenced by the nonvanishing field strength in other sites of the lattice. All the topological states appear within the bulk band gap, with some overlap due to the finite bandwidth of the corresponding spectral lines caused by their finite lifetime. Some overlap among the bulk, surface, edge, and corner spectra, can be seen in Fig.3B and is attributed to inhomogeneous broadening due to energy dissipation. Two major loss mechanisms in our experiment are (i) the absorption in the resin used in 3D printing, (ii) the leakage through the probe holes, which are deliberately introduced in the design of the individual resonators to allow excitation and probing of the acoustic field. The resultant experimentally measured quality factors of the individual resonators are within the range of 50 to 60, and, therefore, the corresponding finite lifetimes of the modes of the lattice are long enough not to alter their topological nature.

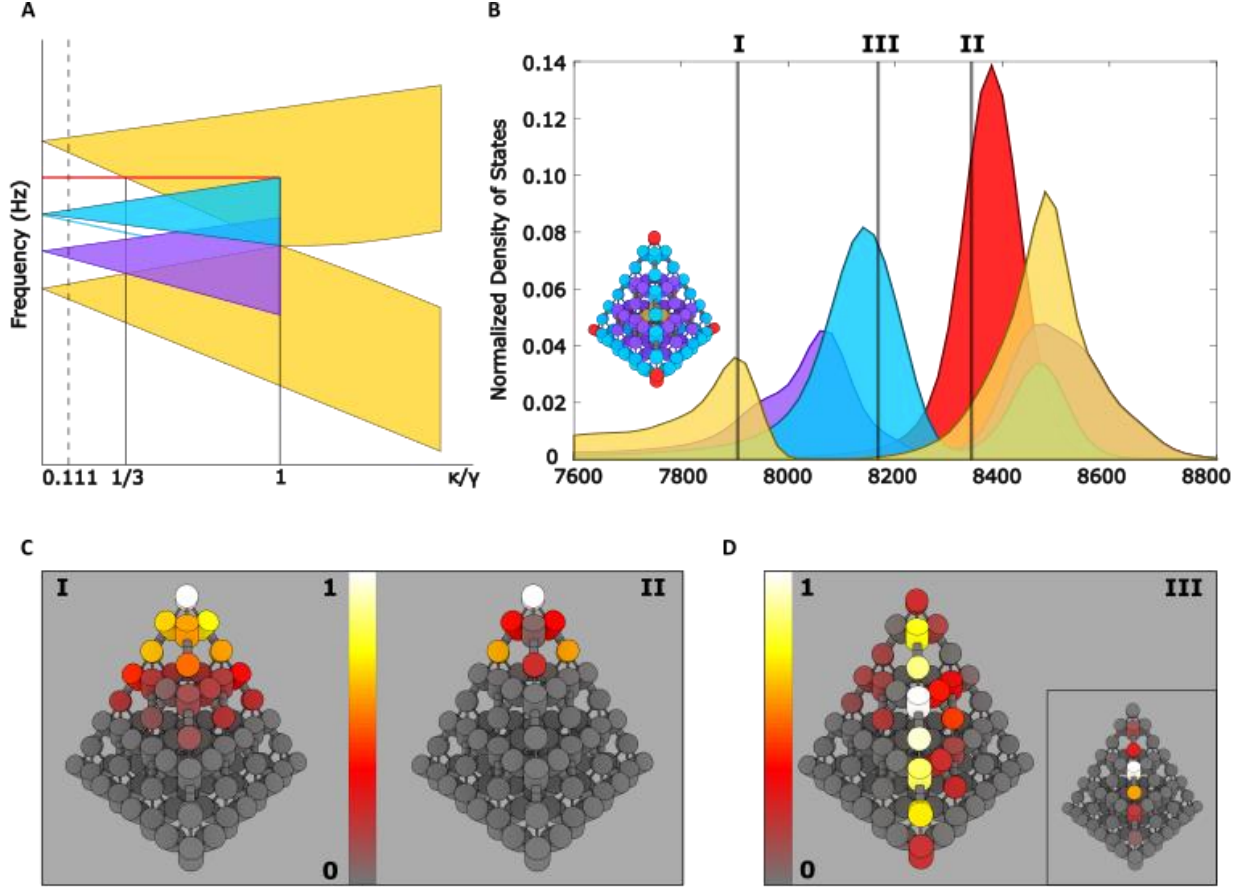

**Fig. 3 Experimental demonstration of second- and third-order topological edge and corner states.** (**A**) Energy spectrum obtained from TBM for the pyrochlore lattice. Yellow, purple, blue, and red colored shaded regions are for bulk, surface, edge, and corner states, respectively. (**B**) Normalized density of states for the expanded lattice obtained from the measurements of the frequency response at the top of each site. Each region's densities of states (bulk, surface, edge, and corner) were obtained by filter functions of bulk, surface, edge and corner, respectively. (**C**) Logarithmic map of the spatial distribution of acoustic power with source placed at the top corner indicated by white colored cylinder evaluated at frequency I and II. (**D**) Logarithmic map of the spatial distributions of acoustic power with source placed at the center edge site (depicted in white colored cylinder) evaluated at frequency III. The inset of (**D**) shows the linear map for the same measurement.

The experimentally measured field profiles of all the modes, shown in Fig. 3C and 3D, are in excellent agreement with the numerically calculated profiles of higher-order states. From Fig. 3C, one can easily ascertain the difference between field profiles of the exponentially decaying bulk modes (excited at the frequency within the bulk band indexed by the label I in Fig. 3B) and of the extremely localized zero-energy corner state excited at the corresponding frequency (indexed by the label II in Fig. 3B). Note that the field decay of the spectral profile of bulk modes is less rapid, following a $1/r^2$ dependence with an exponential correction due to loss. It is worth mentioning that the generalized chiral symmetry in the pyrochlore lattice provides inherent robustness of the corner modes, as the corner states are always pinned to "zero-energy" states, and their field profiles, are ideally localized at one of the sublattices (*19*), which is evident in the corner mode spectral profile in Fig. 3C. The generalized chiral symmetry for the present case of 3D pyrochlore lattice is described by the operator $\hat{\Gamma}_4$, with the property $\hat{\Gamma}_4^4 = 1$, connecting a sequence of inequivalent Hamiltonians $\hat{H}_n$ with the same spectrum $\hat{\Gamma}^n \hat{H}_0 \hat{\Gamma}^{-n} = \hat{H}_n$, where $n = 1,2,3$, and $H_0 + H_1 + H_2 + H_3 = 0$ (S.I. section S3). Indeed, as can be seen from the logarithmic field maps in Fig. 3C, right panel, the corner state predominantly localizes on the same sublattice. Thus, the amplitude of at the atoms of different sublattices within the same unit cell appears to be at least 500 times smaller compared to the amplitude in the same sublattice of the adjacent cells (only 40 time weaker). As expected, the similar sublattice localization is not observed for the bulk modes plotted in the left panel of Fig. 3C.

In addition to corner states localized in three dimensions, the nonzero value of bulk topological polarization gives rise to the formation of another class of HOT states – 2nd order edge states confined to the edges of the fabricated finite metacrystal. These modes are shown in Fig. 3D, where we can see that they exhibit extreme localization to a particular edge, where the mode is excited halfway between the terminating corners, and it forms a standing wave-like profile with two nodes located at the respective corners.

**Conclusions**

Advanced manufacturing techniques based on 3D printing meta-molecules and their snap-assembly to form complex acoustic metamaterials provides exciting opportunities in acoustics. Here, this platform has been applied to fabricate a 3D topological metamaterial forming the acoustic analogue of pyrochlore crystal, and supporting a regime with nontrivial topological bulk polarization. We demonstrated for the first time that such topological polarization leads to the emergence of a hierarchy of higher-order topological states, $3^{nd}$ order states confined to corners, $2^{nd}$ order states localized at the edges of the 3D metacrystal. The respective modes show strong and topologically robust confinement to the respective edges and corners, making them ideal candidates for applications in which precise control of energy localization is critical. In this respect, topological metamaterials not only hold the promise for desirable field concentration, but also add the unique feature of robustness, which ensures the spectral position of the HOT states to be stable, as long as the generalized chiral symmetry of the manufactured structure is ensured by its design and fabrication. The use of additional synthetic degrees of freedom, engineered by the complex geometries of the 3D multimodal metamolecules, may further facilitate control of topological acoustic modes, opening even more interesting applications, with sound-waves filtered, steered, and routed based on their internal synthetic polarization.

## Data availability

Data that are not already included in the paper and/or in the Supplementary Information are available on request from the authors.

## Author contributions

All authors contributed extensively to the work presented in this paper.

## Acknowledgements
The work was supported by the DARPA Nascent program and by the National Science Foundation with grants No. CMMI-1537294, EFRI-1641069, and DMR-1809915. Research carried out in part at the Center for Functional Nanomaterials, Brookhaven National Laboratory, which is supported by the U.S. Department of Energy, Office of Basic Energy Sciences, under Contract No. DE-SC0012704. MW acknowledges the guidance of Benjamin Hahn for the preparation of 3D graphics.

## Competing interests
The authors declare no competing interests.

## Corresponding authors
Correspondence to Andrea Alù (https://orcid.org/0000-0002-4297-5274) or Alexander B. Khanikaev (https://orcid.org/0000-0002-7689-216X).

## Methods

*1. Structure design, 3D printing, and generic measurements* – The unit cell designs of the topological expanded lattice are shown in Fig. 1A-B with lattice constant $a_0 = 61.66$ mm, height $H_0 = 20.00$ mm, and radius $r_0 = 9.70$ mm. The connectors between the cylinders are radial channels with diameter $d_\gamma = 3.66$ mm and $d_\kappa = 1.00$ mm. To preserve the rotational symmetries of the pyrochloric unit cell, we rotated the cylinders about their axis along the z direction by angle $\theta = 18.60°$ in y-z plane (such that the three modes remain degenerate at Γ point, as shown in Fig. 2B). The centers of the four cylinders in each unit cell are aligned to form the FCC lattice. The boundary cells, as shown in the photograph of the finite structure (Fig. 1D), have the same geometric parameters except their heights are 90% that of the cylinders within the unit cell.

The unit cells and boundary cells were fabricated using the B9Creator v1.2 3D printer. All cells were made with acrylic-based light-activated resin, a type of plastic that hardens when exposed to UV light. Each cell was printed with a sufficient thickness to ensure a hard wall boundary condition and narrow probe channels were intentionally introduced on top and bottom of each cylinders to excite and measure local pressure amplitude at each site. The diameter of each port is 2.50 mm with a height of 2.20 mm. When not in use, the probe channels were sealed with plumber's putty. Each unit cell and boundary cell were printed one at a time and the models were designed specifically to interlock tightly with each other. The expanded structure shown in Fig. 1D contains 20 unit cells and 60 boundary cells. For all measurements, a frequency generator and FFT spectrum analyzer scripted in LabVIEW were used.

*2. Numerical method* – Finite element solver COMSOL Multiphysics 5.2a with the Acoustic module was used to perform full-wave simulation. In the acoustic propagation wave equation, the speed of sound was set as 343.2 m/s and density of air as 1.225 kg/m^3. Other dimensional parameters of the structure are the same as the fabricated parameters.

TBM is used to fit the band diagrams of the dimerized and un-dimerized structures with the first principles COMSOL solution. For the un-dimerized band structure with X-W line degeneracy, the on-site frequency is fitted as $\omega_0 = 8584$ Hz and $\kappa = \gamma = 12.8$ Hz. For the dimerized band structure with X-W line broken degeneracy, $\omega_0 = 8470$ Hz and $\kappa = 12.8$ Hz, $\gamma = 115$ Hz ($\kappa$ and $\gamma$ are flipped for the shrunken case). These numbers were used to extract the bulk polarization of the expanded and shrunken cases.

*3. Density of States (DOS) measurement* – The speaker was placed at the bottom port and the microphone at the top port of the same site. The frequency generator was used to run a sweep from 7600 Hz to 8800 Hz in 20 Hz intervals with a dwell time of 0.5 seconds while the FFT spectrum analyzer obtained the amplitude responses φ(ω) at each frequency. Field distributions φⱼ(ω) are obtained by repeating this process for each site *j*. Once each amplitude response was obtained for each of the 80 individual sites, we separate the response of the bulk, surface, edge, and corner sites depending on which mode dominates at that site, put it another word, we use the filter functions to obtain the spectra for different type of modes. We obtained the average power spectrum $P_a(\omega) = \sum_j |\Phi_j(\omega)|^2 /N$ for each region where N is the number of resonators within that region and then normalized the spectra with $P_n(\omega) = P_a(\omega)/\sum_\omega P_a(\omega)$. For the field profiles excited by a single frequency, the speaker was fixed at the port of the site of interest and the microphone was placed over each site of the lattice to measure the magnitude response at the desired frequency (shown in Fig. 3B).